# POROMECHANICAL SOLUTION FOR ONE-DIMENSIONAL LARGE STRAIN CONSOLIDATION OF MODIFIED CAM CLAY SOIL

**SHENG-LI CHEN**
Associate Professor
Department of Civil & Environmental Engineering
Louisiana State University
Baton Rouge, LA 70803, USA
Email: shenglichen@lsu.edu

**HAI-SUI YU**
Professor, FREng
School of Civil Engineering
University of Leeds, Leeds LS2 9JT, UK
Email: DVC@leeds.ac.uk

**YOUNANE N. ABOUSLEIMAN**
Professor Emeritus
School of Geosciences
The University of Oklahoma
Norman, OK 73019, USA
Email: yabousle@ou.edu

**CHRISTOPHER E. KEES**
Associate Professor
Department of Civil & Environmental Engineering
Louisiana State University
Baton Rouge, LA 70803, USA
Email: cekees@lsu.edu

**Abstract:** A theoretical model describing the one-dimensional large strain consolidation of the modified Cam Clay soil is presented in this paper. The model is based on the Lagrangian formulation, and is capable of featuring the variability of soil compressibility (inherently so due to the direct incorporation of the specific Cam Clay plasticity model) and permeability, as well as the impact of overconsolidation ratio. The derivation starts from the establishment of the incremental stress-strain relations for both purely elastic and elastoplastic deformations under one-dimensional compression condition, and thereafter the coefficients of compressibility/volume change that are essential to the consolidation analysis. The governing partial differential equation is then neatly deduced in conjunction with the continuity and equilibrium conditions for the soil, with the vertical effective stress being the privileged unknown to be solved for. Subsequently, semi-analytical solution to the developed rigorous poroelastoplastic large strain consolidation model is obtained and verified with the ABAQUS finite element numerical results. Parametric analyses are finally provided to investigate in detail the influences of the soil overconsolidation ratio, large strain configuration, and the variability of the soil permeability on the calculated one-dimensional consolidation response.

**Introduction**

One of the major simplifying assumptions in Terzaghi's classical one-dimensional theory of consolidation (Terzaghi, 1943; Taylor, 1948) is that the stress-strain relationship remains linear during the consolidation process. Such a simplification/limitation was later removed by numerous researchers to take more realistic account of the nonlinear variation of the soil compressibility during consolidation, within both the small strain (Davis & Raymond, 1965; Poskitt, 1969) and large strain (Gibson et al., 1967; 1981; Carter et al., 1979; Lee & Sills, 1979; Cargill, 1982; Geng & Yu, 2017) theoretical frameworks. A comprehensive review of the one-dimensional infinitesimal/finite strain and linear/nonlinear consolidation theories can be found in Schiffman (2001). Most recently, the Terzaghi consolidation theory has been further extended by Ding et al. (2022) with the inclusion of the inertial effects of the soil mass.

Despite the extensive research on the non-linear consolidation of saturated clay with variable compressibility in the literature, the nonlinearity of the soil behaviour has basically been treated through the use of a relation between the void ratio and soil skeleton stress usually determined experimentally. In this regard, the aforementioned analytical/semi-analytical solutions appear to be of insufficient generality. They indeed lack direct connection with the elastoplastic behaviour of soils (and hence with the well-developed soil plasticity models like Mohr-Coulomb and Cam Clay), which are deemed as the major contributing source to the nonlinear stress-strain relationship, especially in the large strain analysis. An exception is the analysis by Pariseau (1999), who made an important contribution to the analytical modelling of one-dimensional poroelastic-plastic consolidation problem [see Selvadurai (2021)] where the soil skeleton is assumed to obey the ideal Drucker-Prager or Mohr-Coulomb yielding condition/failure criterion. With the adoption of such elastic-perfectly plastic models, the formulation of the one-dimensional small strain elastoplastic consolidation was found to be equivalent to solving a conventional Stefan moving boundary problem, for which the governing diffusion equations take essentially the same form, but with different constant diffusion coefficients, in both the elastic and elastoplastic domains involved (Pariseau, 1999). Recently, Selvadurai (2021) has made numerical efforts to further accommodate the (plastic) irreversibility of the soil skeletal deformations during loading and unloading cycles into the one-dimensional consolidation analysis, by the use of ABAQUS finite element code and an appropriate selection of the elastoplastic constitutive model.

For the consolidation problem of saturated clays, it is natural to employ the Cam Clay critical state models to describe the elastoplastic behaviour of cohesive soil (e.g., Britto, 2013; Selvadurai, 2021). This paper therefore aims to propose a rigorous theoretical formulation and develop subsequently a semi-analytical solution for the one-dimensional large strain consolidation problem, based on the classical modified Cam Clay (MCC) plasticity model. The consolidation model is capable of featuring the variability of soil compressibility and permeability, as well as the due account of soil overconsolidation ratio in a theoretically consistent way. It is recognized that the Cam Clay-type models were initially developed under the triaxial condition and later generalized to the three-dimensional loading state. However, according to Muir Wood (1990: p. 317), the modified Cam Clay model can still reasonably capture the one-dimensional compression response ($K_0$ coefficient) of soil for typical values of Poisson's ratio and plastic volumetric strain ratio encountered. It was recently also found in Hu et al. (2018) that the derived formula for the earth pressure coefficient based on the MCC model is capable of predicting the general variation trend of the measured $K_0$ values during one-dimensional compression. Similar work has been reported by Riad & Zhang (2019) as well for predicting the varying earth pressure coefficient $K_0$ with the vertical effective stress, again with the use of the MCC model. It is worth mentioning that an earlier description of the suitability of applying the modified Cam Clay for numerically modelling the one-dimensional sedimentation and compaction processes at basin scale was provided by Luo et al. (1998).

The formulations start from the derivation of the incremental stress-strain relations, and therefore the coefficients of compressibility/volume change, under one-dimensional compression. These compressibility properties, however, need to include both elastic and elastoplastic scenarios since the former one will be necessarily required for the case of overconsolidated soils. It is shown that when the soil is undergoing purely elastic compression/deformation (which applies only when $OCR > 1$), there exists a completely explicit expression for the variation of void ratio with the vertical effective stress. The elastic and elastoplastic void ratio-effective stress relationships obtained are then combined with the Lagrangian forms of the continuity and equilibrium equations to finally yield a nonlinear differential equation for the MCC one-dimensional large strain consolidation problem. The governing equation is derived with the vertical effective stress being the privileged variable, in a somewhat more straightforward/simple manner compared with the one presented in Gibson et al. (1967). The proposed poroelastoplastic large strain consolidation theory/model, after validation with the ABAQUS (2022) finite element numerical results, is utilized through an example analysis to evaluate the variations of the degree of consolidation with depth as well as the average degree of consolidation settlement of a saturated clay layer. Especially, the impacts of large strain configuration/formulation, the variation of soil permeability, and the overconsolidation ratio on the one-dimensional consolidation progress will be examined in some detail.

It is noteworthy to mention that while the elastoplastic one-dimensional large strain consolidation analysis might be readily performed numerically by using the finite element method (e.g., Britto, 2013; Selvadurai, 2021), the semi-analytical method/solution developed in the present paper, which is believed to be the first one of its kind in the literature, nevertheless could offer the following advantages. On one hand, it is computationally more efficient, and more accessible to a wide range of users (Russell et al., 2023), through solving one single governing differential equation in lieu of resorting to the commercial finite element programs. On the other hand, it can facilitate and provide clearer understanding of the physics and mechanism underlying the consolidation problem via the introduction of pertinent dimensionless parameters and variables controlling the boundary value solution. Furthermore, the new rigorous and accurate semi-analytical solution can be regarded as a unique benchmark for testing the validity of finite element numerical codes involving the large strain coupled poroplasticity.

**One-dimensional compressibility of modified Cam Clay soil**

Consider the one-dimensional large strain consolidation of a modified Cam Clay soil layer resting on a rigid (fixed) base, see Fig. 1. The layer has an initial thickness of $H_0$ and consolidates under an instantaneously applied surface load $q_0$ yet neglecting the effects of its own self weight. Let $\sigma'_v$, $\sigma'_h$, and $u_0$ denote, respectively, the in situ vertical and horizontal effective stresses, and initial (excess) pore water pressure. Following Gibson et al. (1967), it is assumed that at certain time $t$, a soil skeleton element, originally occupying the area $A_0 B_0 C_0 D_0$ bounded by two planes at elevations $a$ and $a + \delta a$ (Fig. 1a), has already moved down and deformed to the current location $ABCD$ with coordinate positions $\xi(a, t)$ and $\xi(a + \delta a, t)$ [Fig. 1b]. Here $a$ and $\xi$ represent the Lagrangian and Eulerian coordinates, respectively (Gibson et al., 1967).

The compressibility of the soil under lateral restraint condition, or the void ratio-vertical effective stress relation, plays a central role in the modelling of one-dimensional consolidation problem. For the current modified Cam Clay soil under consideration, this compressibility property apparently should vary with the (vertical) stress state, which actually can be rigorously derived from the elastoplastic stress-strain relation of the MCC model. However, since the coefficients of soil compressibility take different forms during the purely elastic and elastoplastic consolidation stages, the derivation for the one-dimensional (oedometric) compressibility of the MCC soil will be presented separately for these two distinct deformation phases, as described below.

### ELASTIC DEFORMATION PHASE OF CONSOLIDATION

For an overconsolidated case, the deformations of the soil will remain purely elastic until its stress state reaches the initial yield surface. The elastic stress-strain relationship pertaining to the modified Cam Clay model can be expressed in an incremental form as

$$\begin{Bmatrix} D\varepsilon_x \\ D\varepsilon_z \end{Bmatrix} = \frac{1}{E}\begin{bmatrix} 1-\mu & -\mu \\ -2\mu & 1 \end{bmatrix} \cdot \begin{Bmatrix} D\sigma'_x \\ D\sigma'_z \end{Bmatrix} \tag{1}$$

where $D\varepsilon_x$, $D\varepsilon_z$ and $D\sigma'_x$, $D\sigma'_z$ denote the (elastic) strain increments and effective stress increments in $x$ (horizontal) and $z$ (vertical) directions, respectively; $\mu$ is the Poisson's ratio; and

$$E = \frac{3(1-2\mu)vp'}{\kappa} \tag{2}$$

is the Young's modulus, where $v$ is the specific volume, $p' = \frac{1}{3}(2\sigma'_x + \sigma'_z)$ is the mean effective stress, and $\kappa$ the slope of loading-reloading line in $v - \ln p'$ plane.

Application of the lateral restraint condition $D\varepsilon_x = \varepsilon_x = 0$ for the problem of one-dimensional consolidation, to the first row of Eq. (1), yields

$$\frac{D\sigma'_x}{D\sigma'_z} = \frac{\mu}{1-\mu} \tag{3}$$

which basically indicates that the increase in the horizontal effective stress $\sigma_x'$ is proportional to the increase in the vertical stress $\sigma_z'$, despite the fact that the Young's modulus involved with the modified Cam Clay model does not remain constant during the consolidation process but instead varies with $p'$ (Chen & Abousleiman, 2012). Integrating the above equation and taking into account the initial (in situ) stress conditions, one has

$$\sigma_x' = \frac{\mu}{1-\mu}\sigma_z' + \sigma_h' - \frac{\mu}{1-\mu}\sigma_v' \tag{4}$$

Substitution of Eq. (3) now back to Eq. (1) gives

$$\frac{D\varepsilon_z}{D\sigma_z'} = \frac{-2\mu/E}{(1-\mu)/E}\frac{\mu}{E} + \frac{1}{E} = \frac{\kappa(1+\mu)}{(1-\mu)(1+e)}\frac{1}{2\sigma_x'+\sigma_z'} \tag{5}$$

where $e = v - 1$ denotes the void ratio. With the use of Eq. (4), and if the large strain definition

$$D\varepsilon_z = -\frac{De}{1+e} \tag{6}$$

is adopted, it follows from Eq. (5) that

$$\frac{De}{D\sigma_z'} = -\frac{\kappa}{\sigma_z'+A} \tag{7}$$

where $A = \frac{2(1-\mu)}{1+\mu}\sigma_h' - \frac{2\mu}{1+\mu}\sigma_v'$ is a constant. Eq. (7) can be analytically integrated to obtain the void ratio in closed form as:

$$e = e(\sigma_z') = e_i + \ln\frac{\sigma_v'+A}{\sigma_z'+A} \tag{8}$$

where $e_i$ is the initial void ratio corresponding to the undeformed element $A_0B_0C_0D_0$ at time $t = 0$.

Eqs. (7) and (8) determine the desired coefficient of compressibility, $a_v$ $(= -\frac{De}{D\sigma_z'})$, and the variation of $e$ with $\sigma_z'$, both in fully explicit forms. Once $a_v$ and $e$ are derived, the coefficient of volume change (or volume compressibility), defined as $m_v = \frac{a_v}{1+e}$ (Scott, 1994), can be further

calculated as

$$m_v = m_v(\sigma_z') = \frac{\kappa}{(\sigma_z'+A)(1+\ln\frac{\sigma_h'+A}{\sigma_z'+A})} \quad (9)$$

Obviously the above three expressions (7)-(9) are valid for the stress state only up to $\sigma_z' = \sigma_{z,ep}'$. Here $\sigma_{z,ep}'$ corresponds to the elastic-plastic transition state pertaining to the initial yielding of the soil element, which can be determined by substituting Eq. (4) into the following yield function of the modified Cam Clay model (Muir Wood, 1990)

$$F(p',q,p_C') = q^2 - M^2[p'(p_C' - p')] = 0 \quad (10)$$

where $M$ is the slope of critical state line; $q = \sqrt{(\sigma_x' - \sigma_z')^2}$ is the deviatoric stress; and $p_C'$ is the yield pressure under isotropic compression. Note that $p_C'$ is related to the in situ stresses $\sigma_v'$, $\sigma_h'$ and overconsolidation ratio $OCR$ as follows:

$$p_C' = (1 + \frac{q_i^2}{p_i'^2 M^2})p_i'\, OCR \quad (11)$$

where $p_i' = \frac{1}{3}(2\sigma_h' + \sigma_v')$ and $q_i = |\sigma_v' - \sigma_h'|$.

Combining Eqs. (4) and (10), the elastic-plastic transition vertical effective stress $\sigma_{z,ep}'$ can be obtained (from the resultant quadratic equation) as

$$\sigma_{z,ep}' = \frac{-b_1 + \sqrt{b_1^2 - 4a_1 c_1}}{2a_1} \quad (12)$$

where $a_1 = 9\left(\frac{1-2\mu}{1-\mu}\right)^2 + M^2\left(\frac{1+\mu}{1-\mu}\right)^2$; $b_1 = 2M^2 A\left(\frac{1+\mu}{1-\mu}\right)^2 - 9A\frac{(1+\mu)(1-2\mu)}{(1-\mu)^2} - 3p_C' M^2\left(\frac{1+\mu}{1-\mu}\right)$; and $c_1 = \left(\frac{9}{4} + M^2\right)A^2\left(\frac{1+\mu}{1-\mu}\right)^2 - 3p_C' M^2 A\left(\frac{1+\mu}{1-\mu}\right)$, and where only the root with "+" sign has been found valid for the present one-dimensional consolidation analysis pertaining to a step function loading.

## ELASTOPLASTIC DEFORMATION PHASE OF CONSOLIDATION

When sufficient load/stress has been transferred from the pore water to the soil skeleton (i.e., $\sigma'_z \geq \sigma'_{z,ep}$) to cause initial yielding, plastic deformations will occur. In this situation, the elastoplastic constitutive equation for the modified Cam Clay model, following an analogous procedure as in Chen and Abousleiman (2012) for the cylindrical cavity expansion problem, takes the form

$$\begin{Bmatrix} D\varepsilon_x \\ D\varepsilon_z \end{Bmatrix} = \begin{Bmatrix} 0 \\ D\varepsilon_z \end{Bmatrix} = \begin{Bmatrix} D\varepsilon_x^e \\ D\varepsilon_z^e \end{Bmatrix} + \begin{Bmatrix} D\varepsilon_x^p \\ D\varepsilon_z^p \end{Bmatrix} = \begin{bmatrix} \frac{1-\mu}{E} + 2ya_x^2 & -\frac{\mu}{E} + ya_xa_z \\ -\frac{2\mu}{E} + 2ya_xa_z & \frac{1}{E} + ya_z^2 \end{bmatrix} \begin{Bmatrix} D\sigma'_x \\ D\sigma'_z \end{Bmatrix} \qquad (13)$$

where $D\varepsilon_x^e$, $D\varepsilon_z^e$ and $D\varepsilon_x^p$, $D\varepsilon_z^p$ represent the elastic and plastic strain increments in $x$ and $z$ directions, respectively; $y = \frac{\lambda - \kappa}{vp'^3(M^4 - \eta^4)}$, with $\lambda$ denoting the slope of normal compression line in $v - \ln p'$ plane and $\eta = \frac{q}{p'}$ known as the stress ratio; and

$$a_x = \frac{p'(M^2 - \eta^2)}{3} + 3(\sigma'_x - p'), \quad a_z = \frac{p'(M^2 - \eta^2)}{3} + 3(\sigma'_z - p') \qquad (14)$$

Now with the use of the large strain definition $D\varepsilon_z = -\frac{Dv}{v} = -\frac{De}{1+e}$, Eq. (13) is equivalent to the following:

$$\frac{D\sigma'_x}{D\sigma'_z} = f\left(\frac{\sigma'_x}{\sigma'_z}\right) \qquad (15)$$

$$\frac{De}{D\sigma'_z} = (1 + e)g(\sigma'_x, \sigma'_z) \qquad (16)$$

where

$$f\left(\frac{\sigma'_x}{\sigma'_z}\right) = \frac{\mu\kappa/[3(1-2\mu)p'] - (\lambda-\kappa)/[(M^4-\eta^4)p'^3]a_xa_z}{(1-\mu)\kappa/[3(1-2\mu)p'] + 2(\lambda-\kappa)/[(M^4-\eta^4)p'^3]a_x^2} \qquad (17)$$

$$g(\sigma'_x, \sigma'_z) = -\left\{\frac{\kappa}{[3(1-2\mu)p']} + \frac{(\lambda-\kappa)}{[(M^4-\eta^4)p'^3]}a_z^2\right\} + \frac{\left\{\mu\kappa/[3(1-2\mu)p'] - \frac{(\lambda-\kappa)a_xa_z}{(M^4-\eta^4)p'^3}\right\}\left\{\frac{2\mu\kappa}{3(1-2\mu)p'} - \frac{2(\lambda-\kappa)a_xa_z}{(M^4-\eta^4)p'^3}\right\}}{\frac{(1-\mu)\kappa}{[3(1-2\mu)p']} + \frac{2(\lambda-\kappa)}{[(M^4-\eta^4)p'^3]}a_x^2}$$

$$\tag{18}$$

are both explicitly known functions of $\sigma'_x$ and $\sigma'_z$.

It should be mentioned that unlike the purely elastic deformation, during which stage the void ratio $e$ is analytically expressible in terms of the vertical effective stress $\sigma'_z$ (see Eq. (8)), the $e - \sigma'_z$ relation in the plastic phase has to be implicitly sought from Eqs. (15) and (16) above. This can be most efficiently achieved by directly solving these two first-order differential equations simultaneously and numerically, for $\sigma'_x$ and $e$ as functions of a single variable $\sigma'_z$ [see Chen and Abousleiman (2012) for a similar treatment of the equations involved in the elastoplastic cavity expansion analysis]. Note that the initial conditions involved therein can be readily obtained, from Eqs. (4) and (8), as

$$\sigma'_{x,ep} = \sigma'_x(\sigma'_{z,ep}) = \frac{\mu}{1-\mu}\sigma'_{z,ep} + \frac{1+\mu}{2(1-\mu)}A, \quad e_{ep} = e(\sigma'_{z,ep}) = e_i + \ln\frac{\sigma'_v + A}{\sigma'_{z,ep} + A} \tag{19}$$

where $\sigma'_{x,ep}$ and $e_{ep}$ represent the horizontal effective stress and void ratio at $\sigma'_z$ value of $\sigma'_{z,ep}$ corresponding to the elastic-plastic transition stress state. Apparently, once $\sigma'_x(\sigma'_z)$ and $e(\sigma'_z)$ are known, the coefficients of (volume) compressibility, $a_v(\sigma'_z) = -\frac{De}{D\sigma'_z}$ and $m_v(\sigma'_z) = -\frac{1}{1+e}\frac{De}{D\sigma'_z}$ [$= h(\sigma'_z)$, say], may be in turn straightforwardly determined with the aid of Eq. (16).

In summary, Eqs. (7)-(9) and (15)-(18) form the basis for calculations of the void ratio-vertical stress relation and the stress-dependent soil compressibility. The former set of explicit equations are applicable to the elastic scenario when $\sigma'_z$ is less than the threshold value of $\sigma'_{z,ep}$ as given by Eq. (12), i.e., before the onset of yielding; while the latter implicit ones correspond to the plastic straining conditions $\sigma'_z \geq \sigma'_{z,ep}$. For brevity, these two sets of equations can be united into a single one, formally at least, as

$$\frac{De}{D\sigma_z'} = -a_v(\sigma_z') = \begin{cases} -\frac{\kappa}{\sigma_z'+A} & \sigma_v' \leq \sigma_z' < \sigma_{z,ep}' \\ g(\sigma_x'(\sigma_z'),\sigma_z') & \sigma_{z,ep}' \leq \sigma_z' \leq \sigma_v' + q_0 \end{cases} \quad (20)$$

$$e = e(\sigma_z') = \begin{cases} e_i + \ln\frac{\sigma_v'+A}{\sigma_z'+A} & \sigma_v' \leq \sigma_z' < \sigma_{z,ep}' \\ e_{ep} + \int_{\sigma_{z,ep}'}^{\sigma_z'} g(\sigma_x'(\sigma_z'),\sigma_z') d\sigma_z' & \sigma_{z,ep}' \leq \sigma_z' \leq \sigma_v' + q_0 \end{cases} \quad (21)$$

$$m_v = m_v(\sigma_z') = \begin{cases} \frac{\kappa}{(\sigma_z'+A)[1+\ln(\sigma_h'+A)/(\sigma_z'+A)]} & \sigma_v' \leq \sigma_z' < \sigma_{z,ep}' \\ h(\sigma_z') & \sigma_{z,ep}' \leq \sigma_z' \leq \sigma_v' + q_0 \end{cases} \quad (22)$$

where the requirement of the vertical effective stress varying within the range of $\sigma_v' \leq \sigma_z' \leq \sigma_v' + q_0$ is obvious.

It should be remarked that, different from the small strain consolidation analysis (Scott, 1994), the coefficient of volume compressibility, $m_v$, is actually not directly involved in the formulation of the governing consolidation equation for the current large strain case, Eq. (30) below, as will be described in the following section. However, it is for the purpose of completeness that the elastic and elastoplastic expressions of $m_v$ have been provided in the above equation (22).

**One-dimensional large strain elastoplastic consolidation**

The theoretical framework of one-dimensional nonlinear finite strain consolidation in soils was first developed by Gibson et al. (1967). Their governing equations (with the incorporation of the MCC plasticity), nevertheless, may be more straightforwardly and succinctly derived based on a combined Lagrangian/material-Eulerian formulation, as is described below.

Returning now to the representative element of soil $ABCD$ in the current configuration of the consolidating layer, i.e., the Eulerian coordinate as shown in Fig. 1b. If Darcy's law is assumed to be valid, then the net rate of outflow from the element, $\delta q$, can be expressed as (Scott, 1994)

$$\delta q = -v_s(1+e)\frac{\partial}{\partial \xi}\left[k\frac{\partial(u/\gamma_w)}{\partial \xi}\right] \tag{23}$$

where $u$ denotes the excess pore pressure; $\gamma_w$ is the unit weight of water; $v_s$ is the volume of the soil particles occupying the element $ABCD$, naturally assumed to be constant for an incompressible solid phase; and $k$ is the permeability of the soil, which in general can be regarded as a function of the void ratio such that $k = k(e)$.

Assume further that the water is incompressible, the net rate of flow of water $\delta q$ must equal the rate of reduction of the volume of the voids, i.e., $v_s \frac{De}{Dt}$. Therefore,

$$\frac{1}{\gamma_w}\frac{\partial}{\partial \xi}\left[k(e)\frac{\partial u(\xi,t)}{\partial \xi}\right] = \frac{1}{1+e}\frac{De(\xi,t)}{Dt} \tag{24}$$

which is basically the equation of continuity resulting from the consideration for both the solid and fluid phases (Tan & Scott, 1988), or identical to the fluid continuity equation according to Coussy (2004: p. 13, Eq. (1.62)). Note that here "$D$" on the right side represents the material derivative of the void ratio following the designated soil element $ABCD$ that always embraces the same substance of the solid phase, which should be distinguished from the Eulerian derivative "$\partial$" with respect to the local coordinator $\xi$ on the left side.

Now leveraging the well-established relationship between the Lagrangian coordinate $a$ (Fig. 1a) and the Eulerian coordinate $\xi$ (Fig. 1b) for the one-dimensional finite strain consolidation problem (Gibson et al., 1967), i.e., the solid continuity equation (Coussy, 2004):

$$\frac{\delta \xi}{1+e} = \frac{\delta a}{1+e_i} \tag{25}$$

Eq. (24) can be rewritten as

$$\frac{1}{\gamma_w}\frac{1+e_i}{1+e}\frac{\partial}{\partial a}\left[k(e)\frac{1+e_i}{1+e}\frac{\partial u(a,t)}{\partial a}\right] = \frac{1}{1+e}\frac{De[\xi(a,t),t]}{Dt} \tag{26}$$

Obviously $\frac{De[\xi(a,t),t]}{Dt} \equiv \frac{\partial e(a,t)}{\partial t}$, since the Lagrangian coordinate $a$ is independent upon time $t$.

Hence,

$$\frac{\partial}{\partial a}\left[\frac{k(e)}{\gamma_w}\frac{(1+e_i)^2}{1+e}\frac{\partial u(a,t)}{\partial a}\right] = \frac{\partial e(a,t)}{\partial t} \qquad (27)$$

which has been transformed completely to the Lagrangian formulation. Making further use of the equilibrium equation

$$\Delta\sigma_z'(a,t) + u(a,t) \equiv q_0 \qquad (28)$$

where $\Delta\sigma_z'$ denotes the increase of vertical effective stress, one has

$$\frac{\partial u(a,t)}{\partial a} = -\frac{\partial[\Delta\sigma_z'(a,t)]}{\partial a} = -\frac{\partial[\sigma_z'(a,t)-\sigma_v']}{\partial a} = -\frac{\partial[\sigma_z'(a,t)]}{\partial a} \qquad (29)$$

On substitution of this into Eq. (27), it finally follows that

$$\frac{\partial \sigma_z'(a,t)}{\partial t} = \frac{1}{a_v(\sigma_z')}\frac{\partial}{\partial a}\left\{\frac{k[e(\sigma_z')]}{\gamma_w}\frac{(1+e_i)^2}{1+e(\sigma_z')}\frac{\partial \sigma_z'(a,t)}{\partial a}\right\} \qquad (30)$$

This is equivalent to Gibson et al.'s (1967) governing equation with the effect of self-weight neglected (but derived in somewhat different and more concise way in the present paper), except that the vertical effective stress $\sigma_z'$ instead of the void ratio $e$ has nevertheless been adopted as the dependent variable. The reason of adopting $\sigma_z'$ as the preferred variable lies in the fact that such a slightly modified form of equation (30) renders considerable convenience in formulating the relevant initial and boundary conditions, as well as in the effective incorporation of the piecewise expressions for $a_v(\sigma_z')$ and $e(\sigma_z')$ now in relation to the specific plasticity model of MCC [as already defined through Eqs. (20) and (21)]. Note that the volume compressibility coefficient $m_v$ is indeed not directly involved in the above consolidation equation, as indicated previously.

Eq. (30) describes a nonlinear diffusion equation governing the one-dimensional finite strain consolidation of a modified Cam Clay soil, which is essentially a Stefan problem (Hill & Hart, 1986) due to the altered expressions of $a_v$ and $e$ from purely elastic to elastoplastic phases as given in Eqs. (20) and (21). Seeking for analytical solution to such a Stefan-type differential equation mathematically is formidably difficult. Fortunately, it could be relatively easily solved numerically, for example, by using the symbolic computational package Wolfram Mathematica 12 through the general numerical differential equation solver "NDSolve". To facilitate the investigation of the consolidation behaviour of the clay layer, Eq. (30) may be rewritten in a non-dimensional form as

$$\frac{\partial \bar{\sigma}'_z(\bar{a}, T_v)}{\partial T_v} = \frac{1}{\bar{a}_v(\bar{\sigma}'_z)/\bar{a}_{z0}} \frac{\partial}{\partial \bar{a}} \left\{ \bar{k}[e(\bar{\sigma}'_z)] \frac{1+e_i}{1+e(\bar{\sigma}'_z)} \frac{\partial \bar{\sigma}'_z(\bar{a}, T_v)}{\partial \bar{a}} \right\} \tag{31}$$

where $\bar{a} = \frac{a}{H_{dr}}$, $H_{dr}$ is the length of the maximum drainage path; $\bar{\sigma}'_z = \frac{\sigma'_z}{q_0}$; $\bar{a}_v(\bar{\sigma}'_z) = a_v(\sigma'_z)q_0$; $\bar{a}_{z0} = a_{z0}(\sigma'_v)q_0$, with $a_{z0} = \frac{(1+\mu)\kappa}{3(1-\mu)\sigma'_v}$ denoting a compressibility coefficient like quantity yet corresponding to the initial vertical stress $\sigma'_v$; $\bar{k} = \frac{k(e)}{k_i}$, $k_i$ is the initial soil permeability pertaining to the initial void ratio of $e_i$; and $T_v = \frac{k_i(1+e_i)t}{\gamma_w a_{z0} H_{dr}^2}$ is called the (dimensionless) time factor.

For the one-dimensional consolidation problem as shown in Fig. 1, the initial and boundary conditions are as follows:

$$\bar{\sigma}'_z(\bar{a}, T_v = 0) = \bar{\sigma}'_v, \quad \bar{\sigma}'_z(\bar{a} = 0, T_v) = 1 + \bar{\sigma}'_v, \quad \bar{\sigma}'_z(\bar{a} = 2, T_v) = 1 + \bar{\sigma}'_v \tag{32}$$

where $\bar{\sigma}'_v = \frac{\sigma'_v}{q_0}$, and the latter two boundary conditions may be slightly modified as $\bar{\sigma}'_z(\bar{a} = 0, T_v) = \bar{\sigma}'_z(\bar{a} = 2, T_v) = 1 + \bar{\sigma}'_v - e^{-1000T_v}$ in the Mathematica implementation to effectively avoid the discontinuity of $\bar{\sigma}'_z$ with regard to $T_v$ at the boundaries. Note that in writing Eq. (32), both the upper surface $\bar{a} = \frac{H_0}{H_{dr}} = 2$ and the lower surface $\bar{a} = 0$ have been assumed be pervious, i.e., double drainage condition with $H_{dr}$ equal to $\frac{H_0}{2}$. If the soil layer is fully drainable from the upper boundary but rests on an impervious base, the consolidation analysis for such a single drainage case, however, may be most conveniently carried out by directly leveraging the upper half-layer solution already obtained from its double drainage counterpart (Atkinson &

Bransby, 1978).

Once the distribution of the vertical effective stress $\bar{\sigma}'_z(\bar{a}, T_v)$ is obtained by solving Eq. (31) with the imposed initial/boundary conditions Eq. (32), the local degree of consolidation is given by

$$U_z(\bar{a}, T_v) = \frac{\bar{\sigma}'_z(\bar{a}, T_v) - \bar{\sigma}'_z(\bar{a}, T_v=0)}{\bar{\sigma}'_z(\bar{a}, T_v \to \infty) - \bar{\sigma}'_z(\bar{a}, T_v=0)} = \bar{\sigma}'_z(\bar{a}, T_v) - \bar{\sigma}'_v \quad (33)$$

Furthermore, the average degree of settlement of the layer can be calculated as

$$U(T_v) = \frac{\int_0^1 [e_i - e[\bar{\sigma}'_z(\bar{a}, T_v)]] d\bar{a}}{e_i - e_f} \quad (34)$$

where $e[\bar{\sigma}'_z(\bar{a}, T_v)]$, as a function of $\bar{\sigma}'_z$, is again implicitly determinable from Eq. (21); while $e_f = e|_{\bar{\sigma}'_z = 1 + \bar{\sigma}'_v}$ denotes the final void ratio at the end of consolidation.

## Numerical results

In this section, comparisons will be first made with the ABAQUS finite element model results, to verify the foregoing formulations for the Cam Clay one-dimensional large strain consolidation problem and to check the accuracy of the numerical computations. The validated semi-analytical poroelastoplastic solution will then be employed to investigate the influences on the distributions of local degree of consolidation, $U_z(\bar{a}, T_v)$, and the average degree of consolidation, $U(T_v)$, of the soil overconsolidation ratio, large strain configuration/formulation, as well as the variability of the soil permeability.

### COMPARISON WITH ABAQUS FINITE ELEMENT ANALYSIS

To avoid complicating the situation from consideration of the void ratio dependence of the permeability (which can be implemented in ABAQUS by defining $k$ as a tabular function of $e$, however), a constant value of permeability $k \equiv k_i$ will be assumed in the comparison of analytical and FEM numerical results. The parameters adopted for the validation purpose are as follows: $H_0 = 2H_{dr} = 20$ m (doubly drained); in situ effective stresses $\sigma'_h = \sigma'_v = 50$ kPa (i.e., the initial earth pressure coefficient $K_0 = \frac{\sigma'_h}{\sigma'_v} = 1$) and original excess pore pressure $u_0 = 0$; Cam Clay soil properties $e_i = 1.258$, $\lambda = 0.15$, $\kappa = 0.03$, $M = 1.2$, $\mu = 0.278$ [following Chen & Abousleiman (2012) for Boston Blue clay]; $OCR = 2$ with initial yield pressure $p'_C = 100$ kPa; $k = 1.96 \times 10^{-8}$ m/s, $\gamma_w = 9.8$ kN/m³; and the applied surface load $q_0 = 200$ kPa. Note that in ABAQUS modelling the NLGEOM setting has been turned on to take into account the changes in geometry due to large deformation/strain effects as the consolidation analysis progresses.

Fig. 2 plots the calculated excess pore pressure distributions against depth (distance from the bottom of the layer in terms of the undeformed configuration) for different times of $t = 5, 10, 50$, and 100 days, where the solid lines and circular dots represent the current semi-analytical solution and ABAQUS numerical results, respectively. There is a quite close agreement between the theoretical and finite element solutions. The figure also clearly indicates that the elastoplastic zone initiates at the top and bottom surfaces of the layer, and spreads symmetrically towards its center as the pore pressure gradually decays over the time. The elastic-plastic interface always appears at a depth where the transition vertical effective stress, $\sigma'_{z,ep}$, is just being reached, which corresponds to a dissipated excess pore pressure of $u_{ep} = 136$ kPa for the current set of parameters considered (see the vertical dash line in Fig. 2). Note that at some later times of $t = 50$ and 100 days, the calculated excess pore pressure curves are located entirely on the left side of the threshold vertical line, indicating that the soil layer then has already fully entered into the plastic state. As well, Fig. 3 presents a comparison of the variation of the consolidation settlement of the soil layer with time between the semi-analytical [calculated from Eq. (34)] and ABAQUS solutions. The two predicted results are again found to be in excellent agreement. The validity of the presently developed one-dimensional large strain consolidation model for the modified Cam Clay soil and the accuracy of the semi-analytical solution hence are confirmed.

To demonstrate more clearly the alterations of the expressions for the void ratio and the coefficient of compressibility as soil transitions from elastic to elastoplastic deformation phases (see Eqs. (20) and (21) above), Fig. 4 plots the evolutions of $e$ and $a_v$ with the increase of the vertical effective stress from $\sigma'_z = \sigma'_v = 50$ kPa to $\sigma'_z = \sigma'_v + q_0 = 250$ kPa (semi-analytical results provided only). Obviously, the results presented in this figure apply to all the soil elements in the layer, due to the self-similarity nature of the soil response during the one-dimensional consolidation process. As revealed from Fig. 4, while the void ratio $e$ remains continuous at the elastic-plastic transition stress state $\sigma'_{z,ep} = \sigma'_v + q_0 - u_{ep} = 50 + 200 - 136 = 114$ kPa, discontinuity does occur at this very stress level for its derivative with respect to $\sigma'_z$, i.e., the compressibility coefficient $a_v$ appearing in the consolidation equation (30). The lower plastic part of the $e - \sigma'_z$ plot is observed to have an evidently steeper slope (Fig. 4a) and there exists a significant jump of the magnitude of $a_v$ calculated at $\sigma'_{z,ep}$ (Fig. 4b). Bearing this in mind, one may expect that the two curves related to $t = 5$ and 10 days in Fig. 2 are merely weakly continuous at the elastic-plastic interfaces that occur at $u_{ep} = 136$ kPa, and, accordingly, their slopes with the depth would exhibit certain discontinuity at these interfacial locations.

**PARAMETRIC ANALYSES**

Now parametric analyses will be carried out using Eqs. (33) and (34) to illustrate the impacts of $OCR$, the large strain effect, and the variability of permeability on the consolidation behaviour of the soil layer. As in the preceding validation subsection, the constitutive properties of the modified Cam Clay soil considered herein are again those relevant to Boston Blue clay, i.e., $\lambda = 0.15$, $\kappa = 0.03$, $M = 1.2$, $\mu = 0.278$, and $v_{cs} = 2.74$, where $v_{cs}$ is known as the specific volume at unit pressure on the critical state line (Chen & Abousleiman, 2012). Further, the following relationship based on the Kozeny-Carman equation (Scott, 1994) will be adopted to account for the variation of permeability with void ratio:

$$\frac{k(e)}{k_i} = \frac{e^3}{1+e} \frac{1+e_i}{e_i^3} \tag{35}$$

Table 1 summarizes all the parameters used in the numerical analyses, including the three different values of $OCR = 1, 2$, and 5 (corresponding to $K_0 = 0.5, 1$, and 1.5) involved, and the associated initial vertical effective stress $\sigma_v'$/shear modulus $G_0$ as well as the initial and final void ratios $e_i$ and $e_f$. Note that in the table a single value of $e_i = 1.258$ has been given and $\bar{\sigma}_v'$ $(= \frac{\sigma_v'}{q_0})$ been fixed to be $\frac{1}{5}$ for all the case scenarios, while $\sigma_v'$ is calculated from (Chen & Abousleiman, 2012)

$$1 + e_i = v_{cs} + (\lambda - \kappa)\ln 2 - \lambda \ln\left\{OCR\frac{(1+2K_0)\sigma_v'}{3}\left[1 + \frac{9(1-K_0)^2}{(1+2K_0)^2 M^2}\right]\right\}$$

$$+ \kappa \ln\left\{OCR\left[1 + \frac{9(1-K_0)^2}{(1+2K_0)^2 M^2}\right]\right\} \tag{36}$$

The isochrones for the degree of consolidation $U_z(\bar{a}, T_v)$ against the normalized depth $\bar{a} = \frac{a}{H_{dr}}$ (in terms of the initial undeformed configuration), for the three respective $OCR$ values of 1, 2, and 5, are presented in Fig. 5. While all the three plots shown in this figure appear to bear similar variation trends, the consolidation rate would generally be expedited with the increase of the overconsolidation ratio of the soil. This is a result that has been anticipated and can be explained as follows. The larger the value of $OCR$, the later the soil will tend to develop plastic deformation, provided that $q_0$ is large enough to eventually cause plastic yielding when the applied load/stress has been sufficiently transferred to the soil skeleton. Since the compressibility coefficient $a_v$ could increase considerably as the soil behaviour switches from elastic to plastic phase [see Eq. (20) and Fig. 4b], one may then expect a decrease in the magnitude of $\frac{1}{\bar{a}_v(\bar{\sigma}'_z)/\bar{a}_{z0}}$, a consolidation coefficient like quantity appearing in the governing equation (31) that essentially controls the consolidation process. It is therefore not surprising to see that an increase in $OCR$ leads to a delayed development of plastic response and thus a more rapid rate of the soil consolidation.

In addition, the results in Fig. 5b demonstrate that, for a moderate value of $OCR = 2$, two different regions featuring the elastic and elastoplastic deformations may coexist in the layer at earlier times of $T_v \leq 0.1$. Notice again of the local consolidation degree of $U_{z,ep} = 0.256$ marking an end of the purely elastic phase. This is in contrast with the two other cases of $OCR = 1$ and 5. The former one (Fig. 5a) corresponds to a normally consolidated clay, so there will be no elastic zone existing within the layer, as a result of the immediate occurrence of the plastic yielding at the onset of consolidation; while in the latter case of heavily overconsolidated clay, yielding is actually nowhere possible through the layer, since the relatively low magnitude of the applied load $\frac{q_0}{\sigma'_v} = 5$ (as compared with the initial yield surface size of $\frac{p'_c}{\sigma'_v} = 7.3$ in case of $OCR = 5$) will never trigger any plastic straining even by the end of the consolidation process (see Fig. 5c).

The influences of $OCR$ on the calculated average degree of consolidation, $U$, versus the time factor, $T_v$, are next given in Fig. 6. It can again be seen, and more obviously, that an increased value of $OCR$ (or equivalently, the earth pressure coefficient at rest $K_0$) results in a higher rate of settlement of the consolidating layer. This is indeed consistent with the already made observation from Fig. 6 that the degree of consolidation at different depths in general proceeds more rapidly with the increasing overconsolidation ratio.

Further, to check how impactful would be the incorporation of the large strain configuration on the consolidation progress, Fig. 7 provides comparisons with the conventional small strain theory, in terms again of the local degree of consolidation isochrones $U_z(\bar{a}, T_v)$ and of the average consolidation rate $U$. It should be remarked that in the small strain situation, the elastoplastic constitutive equation remains the same expression (13) as for the large strain case; however, the governing consolidation equation (30) now needs to be altered as

$$\frac{\partial \sigma'_z(\xi,t)}{\partial t} = \frac{1+e(\sigma'_z)}{a_v(\sigma'_z)} \frac{\partial}{\partial \xi} \left\{ \frac{k[e(\sigma'_z)]}{\gamma_w} \frac{\partial \sigma'_z(a,t)}{\partial a} \right\} \tag{37}$$

to adapt to the small strain definition of $D\varepsilon_z = -\frac{De}{1+e_i}$ [instead of $D\varepsilon_z = -\frac{De}{1+e}$ as given in Eq. (6)]. Note that in the figure, a variable permeability $k(e)$ assumed following the above relationship (35) has been adopted as well in generating the small strain solution. It appears from Fig. 7a that at earlier times in the consolidation process, there is little difference between the two solutions. However, at later times as $T_v$ increases from 0.2 to 0.8, the difference becomes increasingly noticeable; the use of the small strain theory tends to yield a slower degree of consolidation than would be predicted by the current finite strain theory. Nevertheless, the effect of large strain configuration/formulation on increasing the calculated average consolidation rate of the layer overall seems to be minor, as indicated in Fig. 7b. This is attributed, to a great extent, to the masking effect induced by the fact that the numerator and denominator of the expression for $U$ [see Eq. (34)] are sort of proportionally influenced by the deformation configuration.

Finally, Fig. 8 presents the comparisons between the solutions with and without considering the dependence of the permeability upon the void ratio, for the calculated results of the isochronic distributions of $U_z$ versus $\bar{a}$ and the consolidation rate $U$ against $T_v$, respectively. As observed from the figure, consideration of the variability of permeability (adoption of Kozeny-Carman model herein) predicts a retarded rate, which is clearly to be expected due to the reduction of $k$ with decreasing $e$ during the consolidation. This is particularly evident when the local degree of consolidation is concerned at intermediate and later times (Fig. 8a).

**Conclusions**

This paper proposes a rigorous theoretical formulation and a semi-analytical solution for the one-dimensional large strain consolidation problem using the modified Cam Clay plasticity model. The consolidation model is capable of accounting for the variability of the soil compressibility and permeability as well as the important soil property of overconsolidation ratio in a theoretically consistent way. Especially, the coverage of the latter overconsolidation ratio renders the present model a unique distinction from the existing literature that has been largely developed from the nonlinear theory of Gibson et al. (1967). The key step in the development of the poroelastoplastic large strain model/solution is found to be the derivation of the coefficients of compressibility under one-dimensional compression condition, which is naturally controlled by the specific soil plasticity model involved (Cam Clay in the present context). A nonlinear partial differential equation governing the one-dimensional large strain consolidation problem, in a form analogous to Gibson et al.'s (1967) yet with the vertical effective stress being the basic variable, is subsequently derived and numerically solved to examine the process of consolidation. It is worth mentioning that, with the availability of the analytical/semi-analytical expressions for the desired compressibility coefficients pertaining to the elastic/elastoplastic deformations, the mathematical difficulties arising from the treatment of a moving elastic-plastic interface that relates to solving nonlinear Stefan problem have been greatly reduced in the present study, through the use of numerical method implemented in symbolic computational program.

The validity of the developed one-dimensional large strain consolidation model for the modified Cam Clay soil and the accuracy of the semi-analytical solution are confirmed through comparisons with the ABAQUS numerical results. Parametric analyses for a representative Cam Clay soil (Boston Blue clay) show that the overconsolidation ratio has a significant influence on the consolidation progress, i.e., an increase in $OCR$ leads to a delayed development of plastic response and thus a more rapid rate of the soil consolidation. Consideration of the large strain configuration/deformation and/or neglect of the void ratio dependence of the permeability tend to yield a faster local degree of consolidation, despite that they might have only minor impacts on increasing the average consolidation rate of the whole soil layer. It should be emphasized that the developed analytical approach for the analysis of one-dimensional large strain consolidation problem is general enough, and theoretically applicable to any elastoplastic models provided that the yield/plastic potential surfaces involved are sufficiently smooth and differentiable. Furthermore, the rigorous and exact solution proposed can be regarded as a benchmark for verifying the correctness and capability of the Cam Clay poroplastic constitutive model built in the commercial FEM programs such as ABAQUS and PLAXIS.

## Acknowledgements

The main work presented in this paper was carried out while the first author was on his sabbatical leave at the University of Leeds during the period of March to June 2024. The first author is grateful for the support of the ACS Petroleum Research Fund, American Chemical Society (PRF# 66583-ND9) and the Industrial Ties Research Subprogram of the Louisiana Board of Regents [LEQSF(2019-22)-RD-B-01] for funding this research.

**Notation**

| | |
|---|---|
| — | upper bar representing dimensionless constants or variables |
| $A$ | constant parameter |
| $ABCD$ | deformed soil element having a coordinate position $\xi$ and thickness $d\xi$ |
| $A_0B_0C_0D_0$ | undeformed soil element having a coordinate position $a$ and thickness $da$ |
| $a$ | Lagrangian coordinate |
| $a_1, b_1, c_1$ | constant parameters |
| $a_x, a_z$ | intermediate variables that are explicit functions of the two stress components $\sigma'_x$ and $\sigma'_z$ |
| $a_v$ | coefficient of compressibility |
| $a_{z0}$ | compressibility coefficient like quantity in terms of initial vertical effective stress $\sigma'_v$ |
| $D$ | material derivative |
| $De$ | change of void ratio |
| $Dv$ | change of specific volume |
| $D\varepsilon_x, D\varepsilon_z$ | total strain increments in $x$ and $z$ directions |
| $D\varepsilon_x^e, D\varepsilon_z^e$ | elastic strain increments in $x$ and $z$ directions |
| $D\varepsilon_x^p, D\varepsilon_z^p$ | plastic strain increments in $x$ and $z$ directions |
| $D\sigma'_x, D\sigma'_z$ | effective stress increments in $x$ and $z$ directions |
| $da$ | thickness of the undeformed soil element in Lagrangian coordinates |
| $d\xi$ | thickness of the deformed soil element in Eulerian coordinates |
| $E$ | Young's modulus |
| $e$ | void ratio |
| $e_{ep}$ | void ratio corresponding to the elastic-plastic transition stress state |
| $e_i$ | initial void ratio |
| $e_f$ | final void ratio |
| $F$ | yield function |
| $f$ | explicit functions of $\frac{\sigma'_x}{\sigma'_z}$ |
| $G_0$ | initial shear modulus |
| $g$ | explicit functions of $\sigma'_x$ and $\sigma'_z$ |
| $H_0$ | thickness of soil layer |
| $H_{dr}$ | length of the longest drainage path |
| $h$ | implicit function of $\sigma'_z$, and is equivalent to $m_v$ |
| $k(e)$ | permeability (function of void ratio $e$) |
| $k_i$ | initial soil permeability pertaining to the initial void ratio $e_i$ |
| $K_0$ | lateral earth pressure coefficient at rest |
| $M$ | slope of the critical state line in $p' - q$ plane |
| $m_v$ | coefficient of volume change/compressibility |
| $p'$ | mean effective stress |
| $p'_C$ | yield pressure under isotropic compression |
| $p'_i$ | initial mean effective stress |
| $q$ | deviatoric stress |
| $q_0$ | magnitude of surcharge loading |
| $q_i$ | initial deviatoric stress |
| $T_v$ | dimensionless time factor |
| $t$ | time |
| $U_z$ | local degree of consolidation |
| $U_{z,ep}$ | local degree of consolidation corresponding to the end of pure elastic phase |
| $U$ | average degree of consolidation, or degree of settlement of the entire soil layer |
| $u$ | excess pore water pressure |
| $u_0$ | initial excess pore water pressure |
| $u_{ep}$ | excess pore water pressure at elastic-plastic transition state |
| $v$ | specific volume |
| $v_{cs}$ | specific volume at unit $p'$ on critical state line in $v - \ln p'$ plane |
| $v_s$ | volume of the soil particles occupying the representative element |
| $y$ | intermediate variable |
| $\gamma_w$ | unit weight of water |
| $\Delta\sigma_z'$ | increase in vertical effective stress |
| $\delta q$ | net outflow rate |
| $\eta$ | stress ratio |
| $\kappa$ | slope of the loading-reloading line in $v - \ln p'$ plane |
| $\lambda$ | slope of the normal compression line in $v - \ln p'$ plane |
| $\mu$ | Poisson's ratio |
| $\xi$ | Eulerian coordinate |
| $\sigma_h'$ | in situ horizontal effective stress |
| $\sigma_v'$ | in situ vertical effective stress |
| $\sigma'_x, \sigma'_z$ | effective stress components in $x$ and $z$ directions |
| $\sigma'_{x,ep}$ | horizontal effective stress corresponding to the elastic-plastic transition stress state |
| $\sigma'_{z,ep}$ | elastic-plastic transition vertical effective stress pertaining to the onset of yielding |

**Captions of tables and figures**

Table 1. Soil parameters used for elastoplastic consolidation analysis (Boston Blue clay)

Fig. 1. Consolidating modified Cam Clay soil layer: (a) initial configuration t = 0; (b) current configuration at time t (modified after Gibson et al. (1967))

Fig. 2. Comparison of variations of excess pore water pressure with layer depth (with respect to initial undeformed configuration) between current semi-analytical solution (solid lines) and ABAQUS numerical results (circular dots)

Fig. 3. Comparison of consolidation settlement between semi-analytical and ABAQUS solutions

Fig. 4. Evolutions of (a) void ratio and (b) compressibility coefficient of soil with vertical effective stress

Fig. 5. Isochrones of degree of consolidation against depth: (a) $OCR = 1$ ($K_0 = 0.5$); (b) $OCR = 2$ ($K_0 = 1$); (c) $OCR = 5$ ($K_0 = 1.5$)

Fig. 6. Influences of overconsolidation ratio on variation of average degree of consolidation with time factor

Fig. 7. Impacts of large strain deformation on the process of consolidation: (a) local degree of consolidation isochrones (solid and dashed lines represent large strain and small strain solutions, respectively); (b) average consolidation rate

Fig. 8. Impacts of varying permeability on the process of consolidation: (a) local degree of consolidation isochrones (solid and dashed lines represent variable and constant permeability, respectively); (b) average consolidation rate

**Table 1. Soil parameters used for elastoplastic consolidation analysis (Boston Blue clay)**

| $\lambda = 0.15, \kappa = 0.03, \mu = 0.278, v_{cs} = 2.74, M = 1.2, e_i = 1.258, \bar{\sigma}'_v = \frac{1}{5}$ | | | | | |
|---|---|---|---|---|---|
| OCR | $K_0$ | $\sigma'_v$ (kPa) | $q_0$ (kPa) | $G_0$ (kPa) | $e_f$ |
| 1 | 0.5 | 49.83 | 249.2 | 1303 | 0.992 |
| 2 | 1 | 24.86 | 124.3 | 975 | 1.096 |
| 5 | 1.5 | 8.32 | 41.9 | 435 | 1.223 |

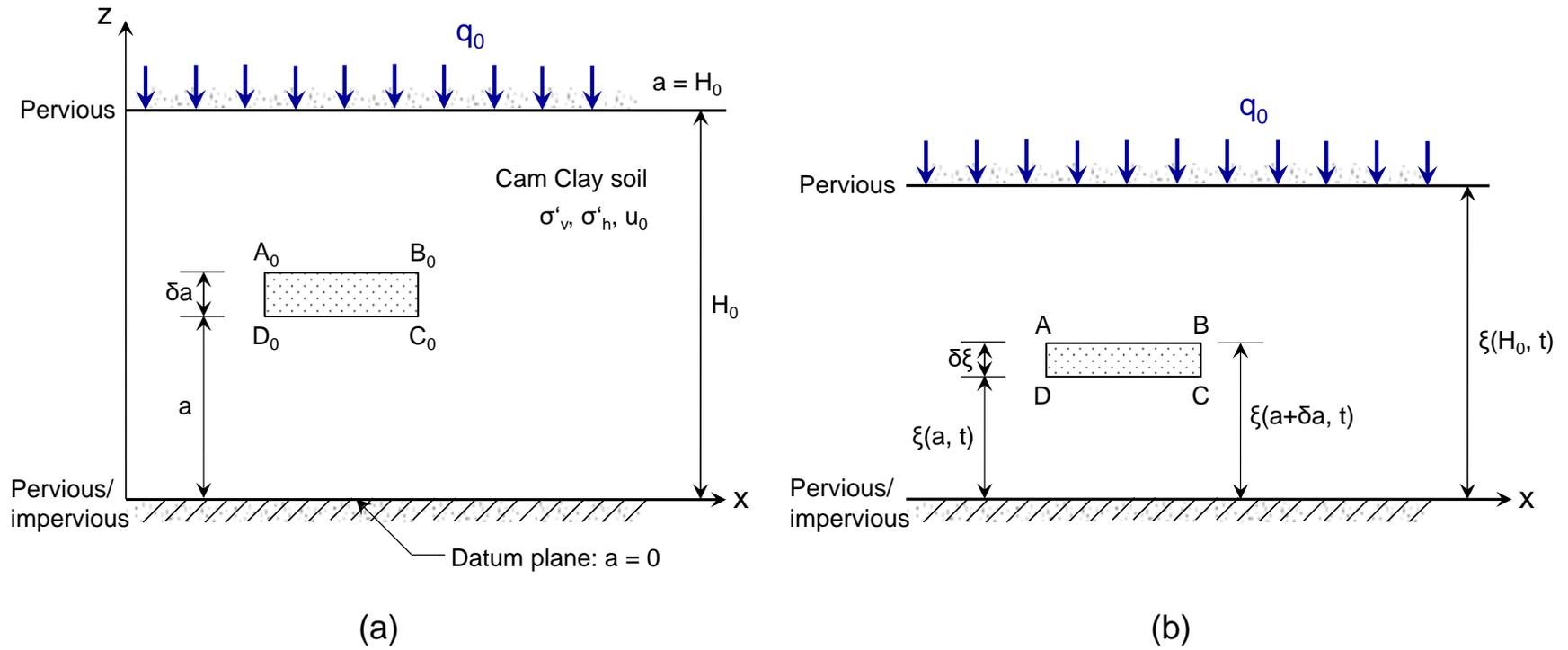

**Fig. 1. Consolidating modified Cam Clay soil layer: (a) initial configuration t = 0; (b) current configuration at time t (modified after Gibson et al. (1967))**

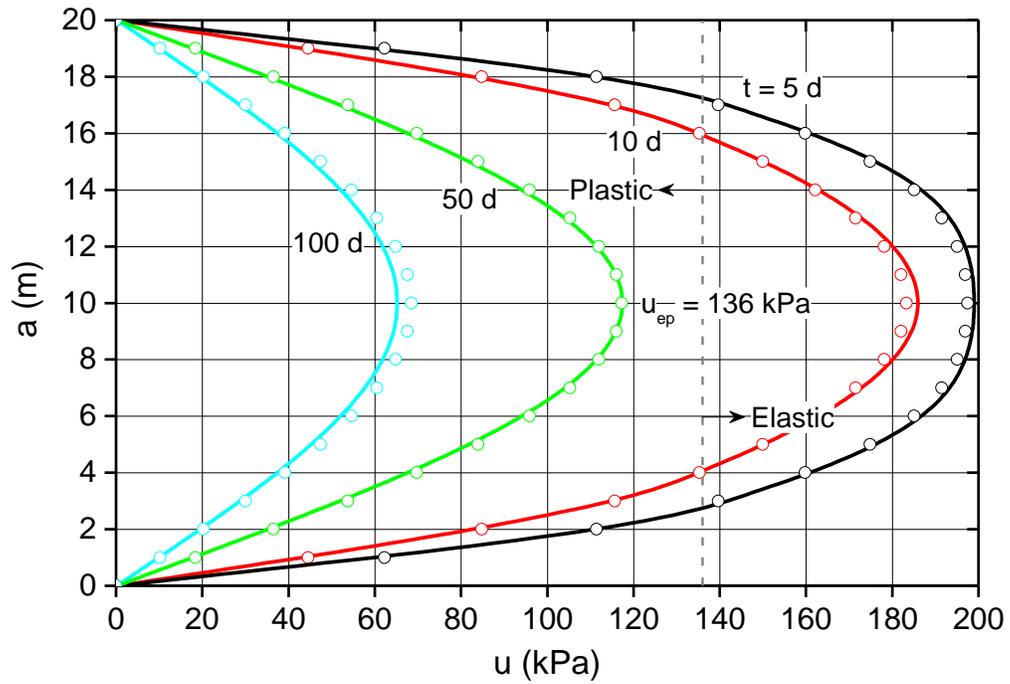

**Fig. 2.** Comparison of variations of excess pore water pressure with layer depth (with respect to initial undeformed configuration) between current semi-analytical solution (solid lines) and ABAQUS numerical results (circular dots)

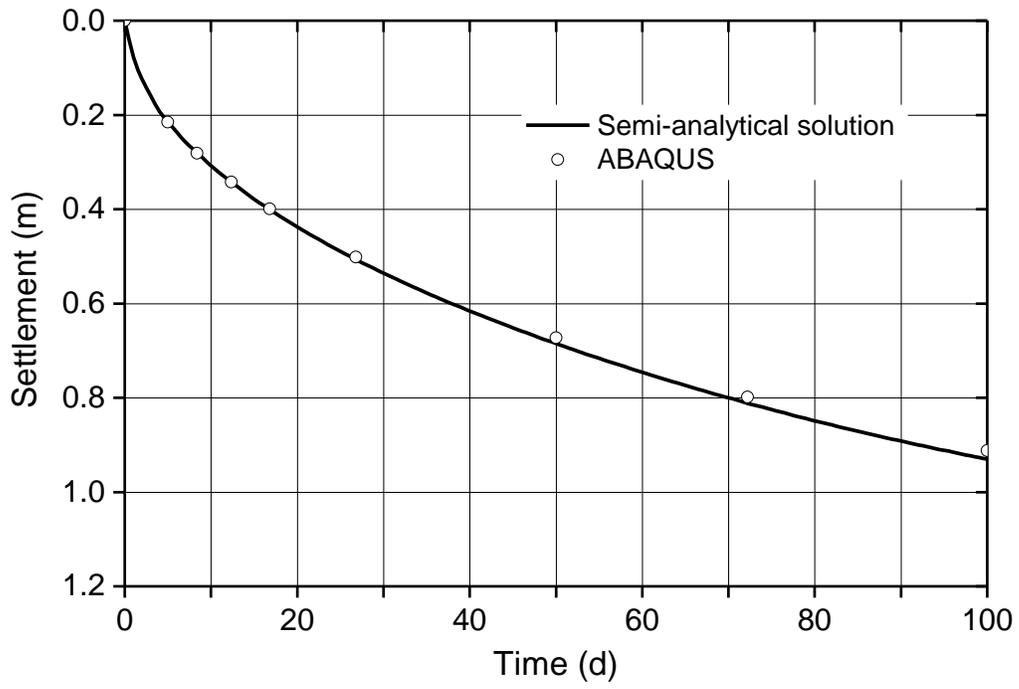

**Fig. 3.** Comparison of consolidation settlement between semi-analytical and ABAQUS solutions

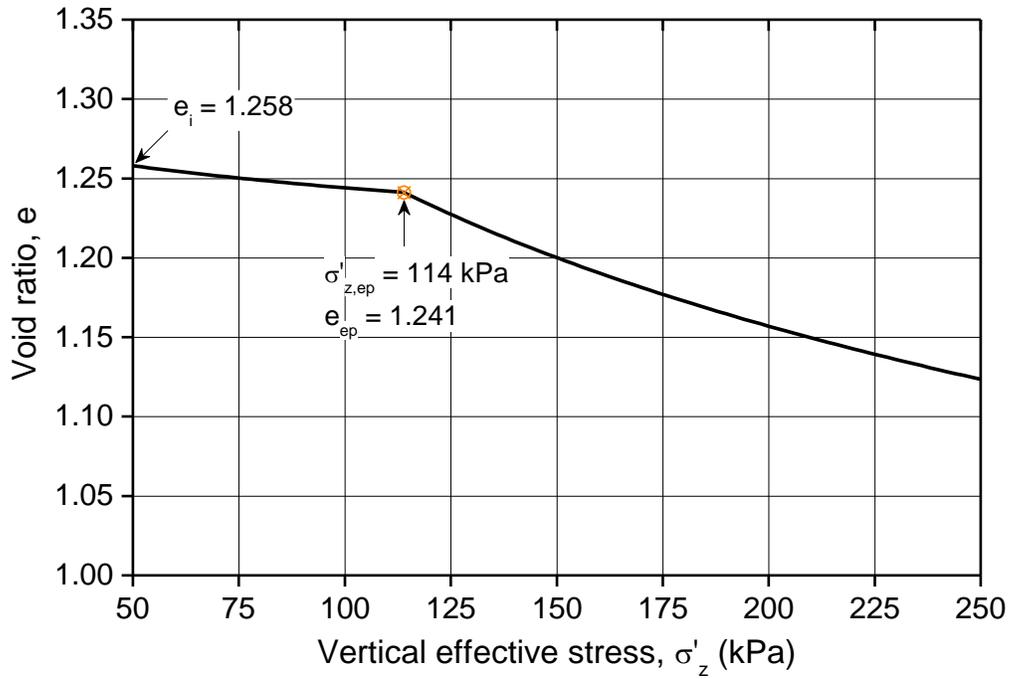

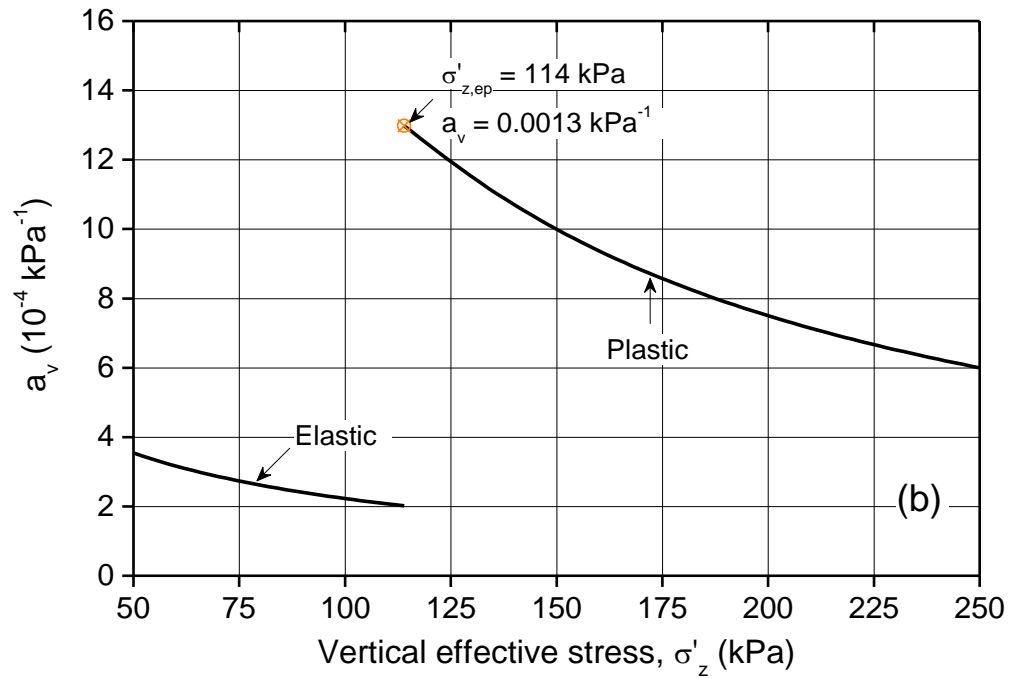

**Fig. 4. Evolutions of (a) void ratio and (b) compressibility coefficient of soil with vertical effective stress**

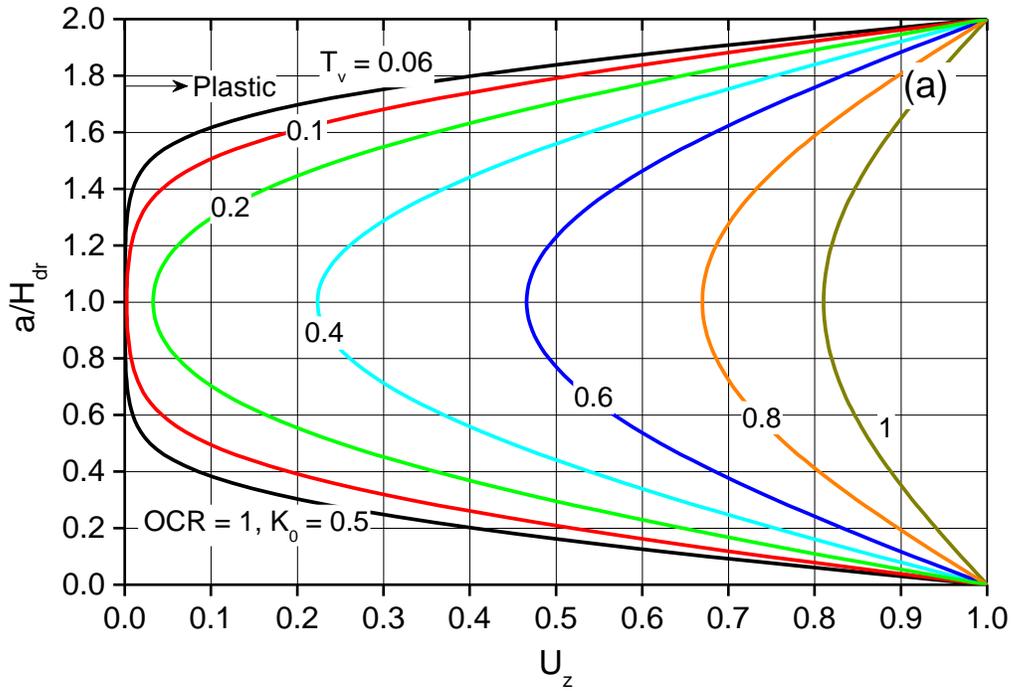

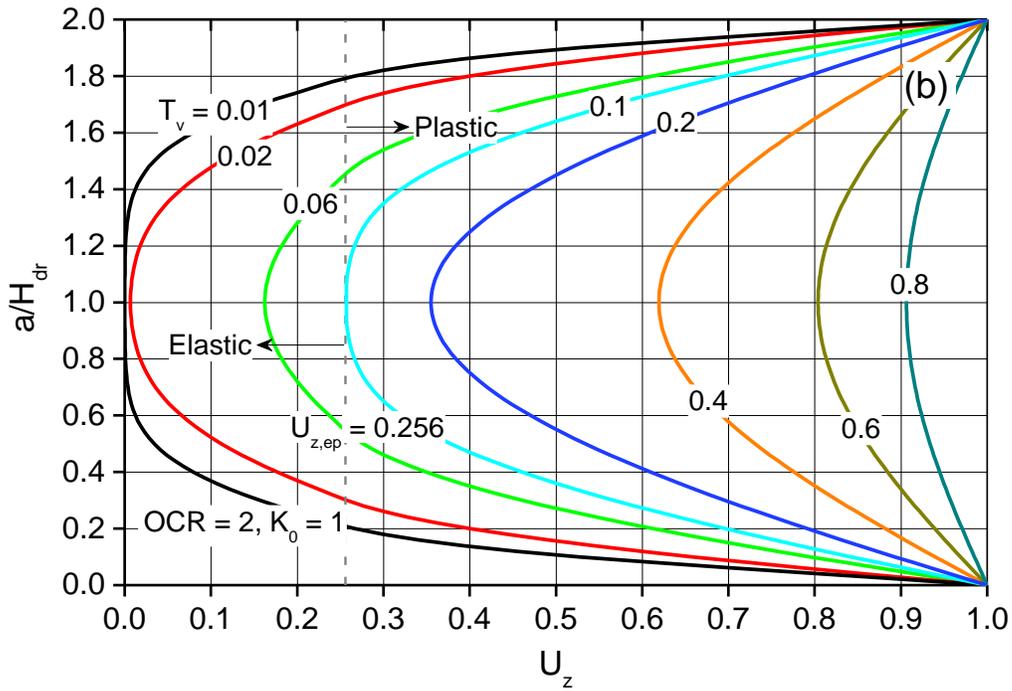

**Fig. 5.** Isochrones of degree of consolidation against depth: (a) OCR = 1 ($K_0 = 0.5$); (b) OCR = 2 ($K_0 = 1$); (c) OCR = 5 ($K_0 = 1.5$)

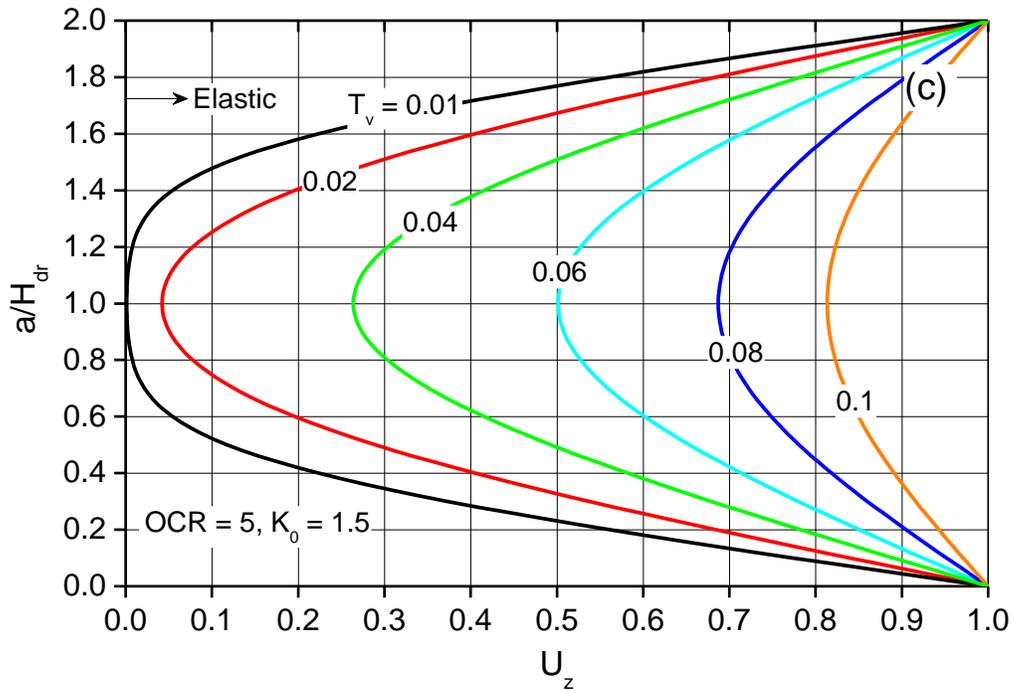

Fig. 5. (Cont'd)  Isochrones of degree of consolidation against depth: (a) OCR = 1 ($K_0$ = 0.5); (b) OCR = 2 ($K_0$ = 1); (c) OCR = 5 ($K_0$ = 1.5)

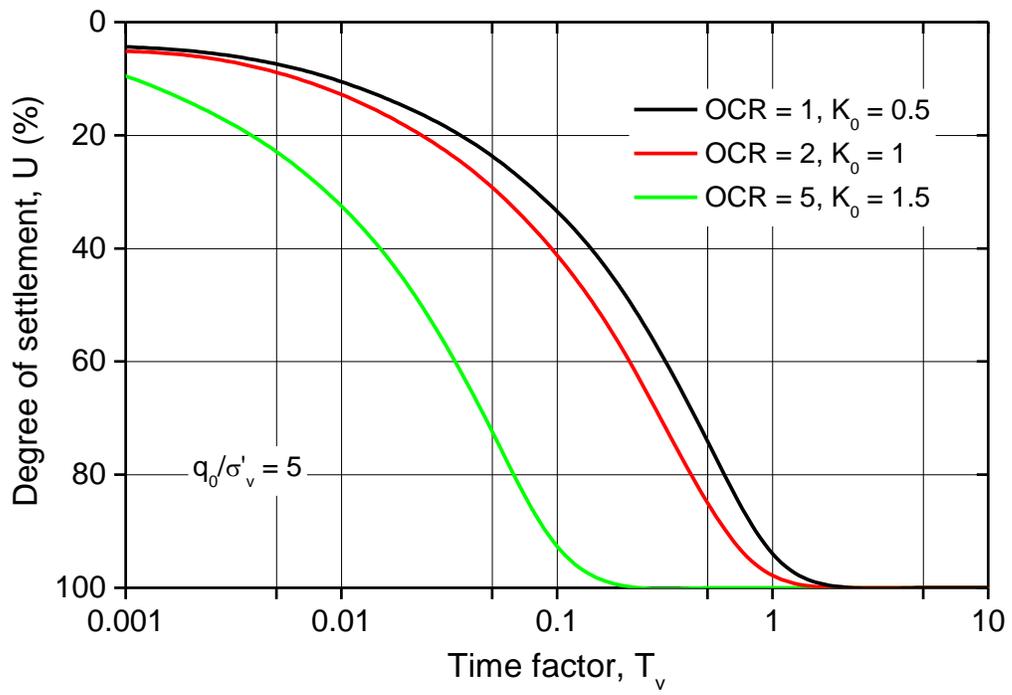

**Fig. 6.** Influences of overconsolidation ratio on variation of average degree of consolidation with time factor

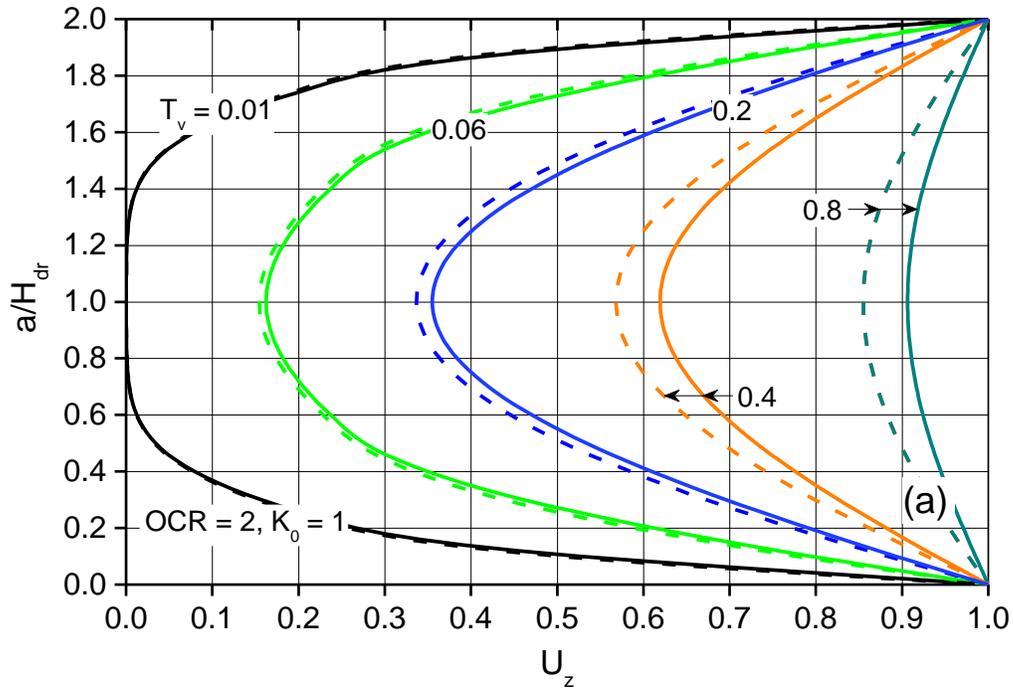

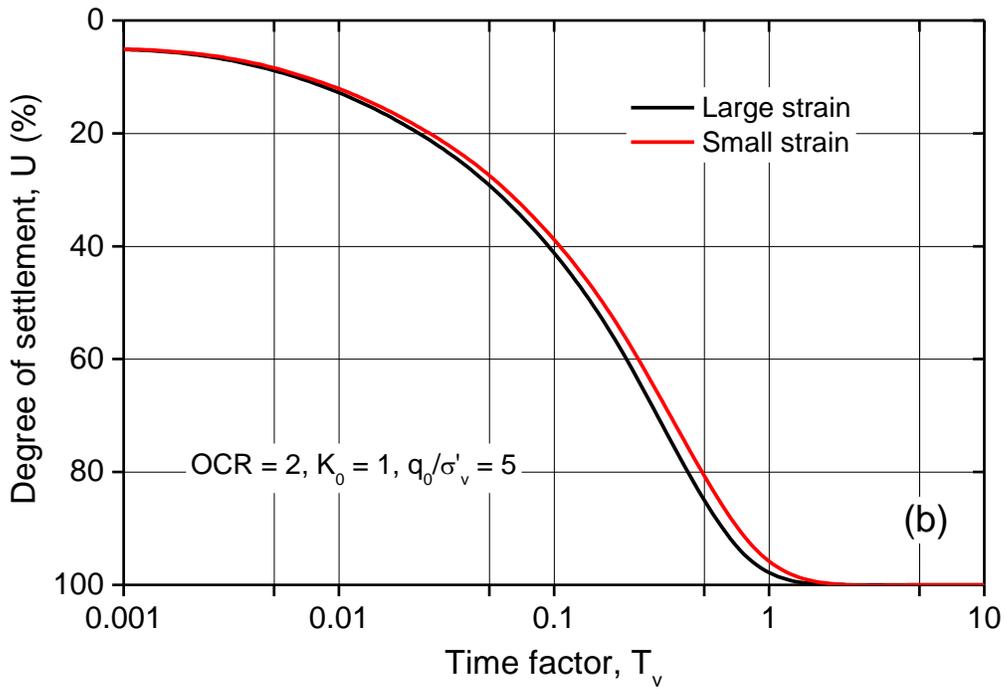

Fig. 7. Impacts of large strain deformation on the process of consolidation: (a) local degree of consolidation isochrones (solid and dashed lines represent large strain and small strain solutions, respectively); (b) average consolidation rate

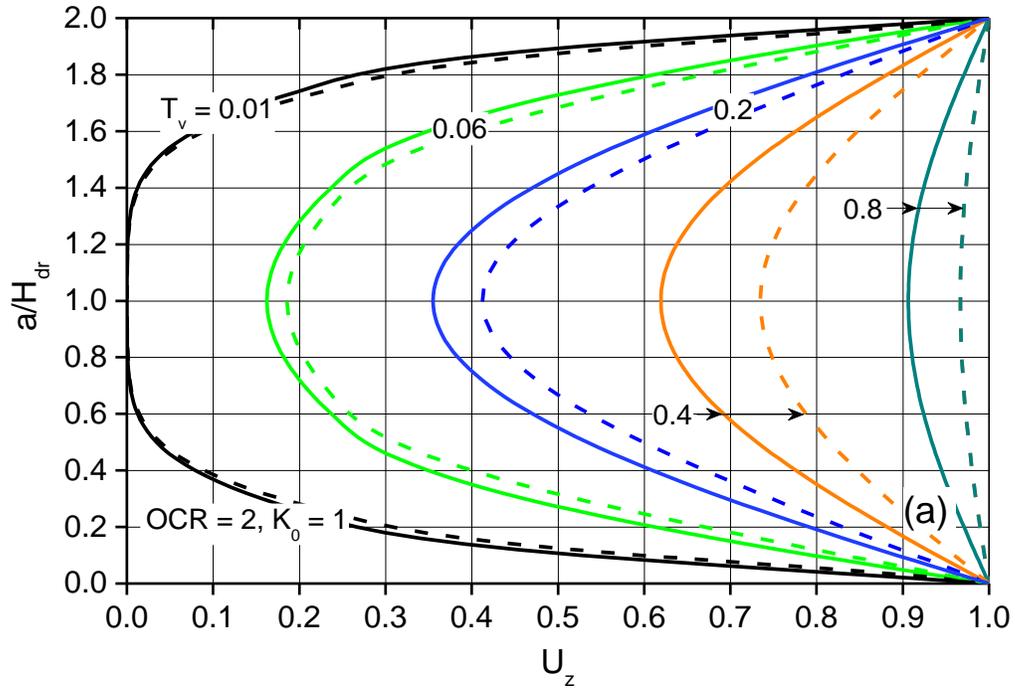

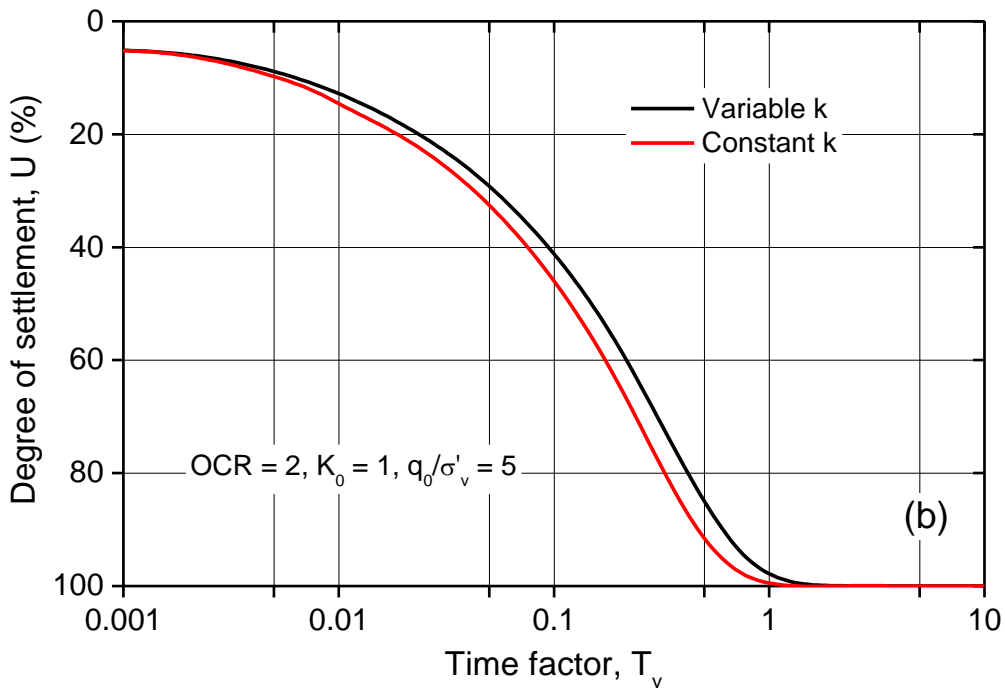

**Fig. 8.** Impacts of varying permeability on the process of consolidation: (a) local degree of consolidation isochrones (solid and dashed lines represent variable and constant permeability, respectively); (b) average consolidation rate